\documentclass{aastex}          
\usepackage{spr-astr-addons}    
\usepackage{graphicx}
\usepackage{latexsym}

\RequirePackage{color}
\newcommand{\bq}       {\begin{eqnarray}}
\newcommand{\eq}       {\end{eqnarray}}
\newcommand{\rf}[1]    {(\ref{#1})}

\newcommand{\bp}       {\begin{minipage}}
\newcommand{\ep}       {\end{minipage}}

\begin{document}
%
\title{Hydrodynamics of Hypersonic Jets: Experiments and Numerical Simulations}

\shorttitle{Hydrodynamics of Hypersonic Jets:}
\shortauthors{M. Belan et al.}

\author{Marco Belan, Sergio de Ponte}
\affil{\it Politecnico di Milano - Dipartimento di Ingegneria Aerospaziale, Milano, Italy}
\author{Daniela Tordella}
\affil{\it Politecnico di Torino - Dipartimento di Ingegneria Aerospaziale, Torino, Italy}
\author{Silvano Massaglia, Attilio Ferrari, Andrea Mignone}
\affil{\it Universit\`a di Torino - Dipartimento di Fisica Generale, Torino, Italy}
\author{Eberhard Bodenschatz}
\affil{\it Max Planck Institute for Dynamics and Self-Organization - G\"ottingen, Germany}

\begin{abstract}

Stars form in regions of the galaxy that are denser and cooler than the mean interstellar
medium. These regions are called Giant Molecular Clouds. At the beginning of their life,
up to $10^5-10^6$ years, stars accrete matter from their rich surrounding environment and are
origin of a peculiar phenomenon that is the jet emission. Jets from Young Stellar Objects (YSOs)
are intensively studied by the astrophysical community by observations at different wavelengths,
analytical and numerical modeling and laboratory experiments. 
Indications about the jet propagation and its resulting morphologies are here obtained by means of a combined study 
of hypersonic jets carried out both in the laboratory and by numerical simulations.

\end{abstract}

\section{Introduction}

The physical parameters of astrophysical jets are impossible to reproduce in any Earth's laboratory.
We therefore must limit ourselves to identify those parameters that are the crucial ones in
controlling the overall jet behavior and to rescale the laboratory simulations accordingly. 
Among the many parameters that describe the jet evolution and propagation features, two 
of them are paramount: the Mach number and the jet to ambient density ratio \citep{ferr98,rb01}. We have
therefore devised an experiment where these two parameters can be set to match the ones 
derived from observations of YSO jets \citep{apss}.

In order to better understand the dynamical details of the jet propagation and evolution, 
we have carried out 3D numerical hydrodynamic simulations adopting boundary and
injection conditions that were as close as possible to the experimental ones. The
numerical results show good agreement with the laboratory data, namely: the jet head
velocity is consistent for the two situations, the presence of bright knots along
the jet is correctly reproduced, and non-axisymmetric perturbations, seen in the heavy jet case,
rise and grow in the simulation to resemble the laboratory jet.

Observations show that stellar jets extend in space for very long distances, 
in fact they can reach lengths
of hundreds, even thousands, jet radii. While the initial collimation, at the launching region, 
is likely to be magnetic, however at very large distances from the origin stellar jets maintain
their small free-jet opening angle, consistently with a Mach number ranging between 10 and 40. 
Here we show that laboratory hydro hypersonic jets maintain their structure, at least up to 70 jet radii.

The plan of the paper is the following: in Section 2 and 3 we present the apparatus setup and the numerical code,
Section 4 is dedicated to the comparison of experimental and numerical results, and the conclusions 
are drawn in Section 5.

\section{Experimental setup}

\begin{figure}[ht]
\centering
\includegraphics[width=\columnwidth]{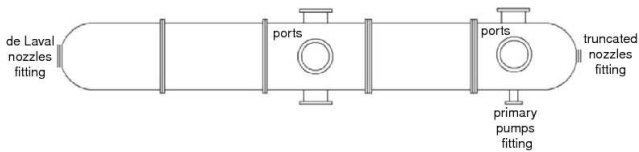}
\caption{Sketch of the vacuum vessel including all the available sections. The ports can be used for camera or electron gun mounting.
Ambient gas valve fittings and electrical/service connectors are not shown.}
\label{vessel}
\end{figure}

The jets under study are created inside a modular vacuum vessel, with a maximum length of 5m and a diameter of 0.5m. 
Several kinds of nozzles can be mounted on this vessel. The vessel diameter is much larger than the diameter of the jets, so that the lateral walls effects are limited. The required ambient inside the vessel is obtained by means of a system of valves which sets the desired ambient density (at pressures in the  1.5 to 100 Pa range) using a gas in general different from the jet gas. An electron gun operating at very low pressures, thanks to a set of secondary pumps, and equipped with a deflection system, creates an electron sheet. This sheet intercepts the jet under test, and generates a plane fluorescent section of the flow. These fluorescent zones can be acquired by different kinds of cameras, including intensified devices, depending on the experiment to be performed. The cameras and the electron gun can be mounted on different ports. The vessel including all the available sections is shown in 
fig. \ref{vessel}.

This system was already employed to study the properties of highly underexpanded jets, making use of truncated 
nozzles and color CCD cameras, detailed information can be found in the previous works by 
\cite{apss,kth,expf}.

In the present setup, by means of a fast piston, the jet gas  is compressed to stagnation pressures
in the  0.1 to 0.7 MPa range, then it is accelerated by a de Laval nozzle and enters the vacuum vessel, 
mounted in the 3-modules configuration. 
The jet parameters attainable by the experimental system are outlined in table \ref{tab}. 

The nozzle used in this experiment has been carefully designed
to perform nearly isoentropical gas accelerations and expansions, i.e. flows as close as possible to the ideal 
conditions, but accounting for friction and heat exchange effects.

The nozzles were built at the CERN workshop of Geneva with surface accuracies
in the order of 1$\mu$m. Under such conditions, the flow at the nozzle outlet has the design Mach number 
over a very wide
cross-section, excluding a thin boundary layer region at the walls, provided that the stagnation-to-ambient 
pressure ratio $p_0/p_a$ maintained between inlet and outlet has a well defined value. In this experiment, 
 the stagnation-to-ambient pressure ratio $p_0/p_a$ is checked by means of high 
accuracy pressure transducers. The Mach number is then obtained by a numerical
method that takes into account the effects of the boundary layer and heat exchange
with the nozzle walls. However the result differs by the well-known steady-state ideal 
relationship:
$$
M^2=\frac{2}{\gamma-1}\left[\left(\frac{p_0}{p_a}\right)^{\frac{\gamma-1}{\gamma}}-1\right]
$$
by about 1\%.

\begin{table}[h]
\small
\caption{Jets parameters in the present experiment}
\label{tab}
\begin{tabular}{@{}ll}
\hline
Nozzle geometry                                 & De Laval \\
Stagnation/ambient pr. ratio                    $p_0 / p_{\rm amb}$ & 300 to $2 \cdot 10^5$ \\
Mach number $M_{max}$                           & 5 to 20 \\
Density ratio $\rho_{\rm jet}/\rho_{\rm amb}$   & 0.01 to 110 \\
Reynolds, throat diameter based $Re_n$          & $10^4$ to $5 \cdot 10^4$ \\
Reynolds, jet diameter based $Re_D$             & up to $2 \cdot 10^6$  \\
Reynolds, axial length based $Re_x$             & $> 10^7$\\
\hline
\end{tabular}
\end{table}

\begin{figure}[h]
\centering
\includegraphics[width=\columnwidth]{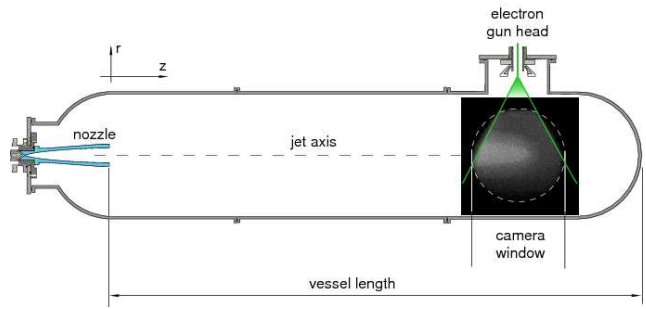}
\caption{Typical setup for matched jets at M=15}
\label{setup}
\end{figure}

Several nozzles in the Mach number range $5 \le M \le 20$ are available,
in this experiment a nozzle designed to generate  a $M$ = 15 at the exit has been used. 
The nozzle exit diameter is $2 r_0=0.07136$ m.
Slightly different Mach numbers can be obtained by adjusting the stagnation/ambient pressure 
ratio $p_0/p_a$ around the ideal isoentropic value $4.76 \cdot 10^4$. 
The present setup is sketched in fig. \ref{setup}, where the vessel has a length of 2.46 m, and the camera 
(optical) window is 0.27 m wide.

In each run a pair of gases is selected for the jet and the ambient. The available gases are helium, argon, 
xenon and air, only the first three
ones are used for the jet since the nozzle is designed for monoatomic gases. 
The jets meet the electron sheet under the optical window, and the resulting images are acquired by an 
intensified CMOS camera;
besides visualizations, density and structure velocities measurements are possible by image-processing.

\begin{figure}[h]
\centering
\includegraphics[width=\columnwidth]{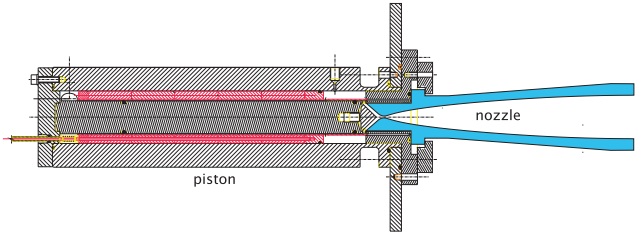}
\caption{ Longitudinal section of the piston-nozzle system. The piston is shown at the starting position}
\label{pist}
\end{figure}

The physical time scale of jets $\tau$ can be defined as the ratio between the jet radius at the nozzle exit 
$r_0$ 
and the speed of sound internal to the jet under nominal conditions, i.e. when the nozzle works under ideal 
steady conditions,
and is in the order of 0.1 to few ms. In the following, we will express time in units of $\tau$ and lengths 
in units of $r_0$.

\begin{figure}[h]
\centering
\includegraphics[width=0.8 \columnwidth]{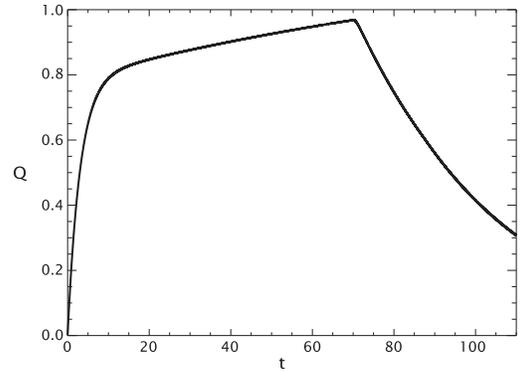}
\caption{A typical jet gas injection: dimensionless nozzle mass flow versus dimensionless time.
The mass flow $Q$ is normalized over the reference flow rate $Q_r$ which is yielded by 
the same nozzle when operating in ideal steady conditions.
For the Helium jet, $Q_r$ is $ 0.43$ g/s and for the Xenon jet, $Q_r$ is $2.44$ g/s.
Time is normalized over the jet time unit $\tau$ (ratio between the jet radius
at the nozzle exit and the speed of sound internal to the jet).
The piston output valves begin to open at $t=0$.
In general, the jet hits the vessel end during the increasing phase of $Q(t)$}
\label{inj}
\end{figure}

The piston-nozzle system is shown in fig. \ref{pist}. The jet gas is compressed in a chamber that leads to the nozzle inlet. 
At the end of the compression run, owing to the valves opening, the outflow increases to a maximum value, 
then it diminishes as the gas contained in the chamber is used up.
The mass flow $Q$ is a function of time of the kind
$$Q = Q [p_0(t), f(t), g(t)],$$
where {$p_0$} is the  pressure in the piston chamber, {$f(t)$} is a factor function which accounts for the 
progressive opening of the valves, and {$g(t)$}  is the function which accounts for
the natural decay in the amount of gas remaining in the compression chamber after the piston stop.
This model, which is based on the piston-valve behavior, gives the following for density, pressure and velocity of the gas at the nozzle exit:
\begin{eqnarray}
\rho_{\rm jet} \sim Q(t)^{2/3}, \; p_{\rm jet} \sim Q(t)^{2/3}/\gamma, v_{\rm jet} \sim { M} \ Q(t)^{1/3} ,
\nonumber
\end{eqnarray}
where  $M$ is the Mach number at the end of the nozzle.
The resulting gas injection curve has the shape sketched in fig. \ref{inj}.
In this experiment two jets are considered, a light He jet flowing in a Xe ambient and a heavy Xe jet flowing in an air ambient.

\section{The numerical code PLUTO}

The numerical code was developed for the solution of hypersonic flows in 1, 2, and 3 spatial dimensions and different systems of coordinates
\citep{mig07,mig09}. The code provides a multiphysics, multialgorithm modular environment particularly
oriented toward the treatment of astrophysical flows in presence of discontinuities. Different hydrodynamic modules
and algorithms may be independently selected to properly describe Newtonian, relativistic, MHD, or relativistic MHD
fluids.

The modular structure exploits a general framework for integrating a system of conservation laws, built on modern
Godunov-type shock-capturing schemes. The discretization recipe involves three general
steps: a piecewise polynomial reconstruction followed by the solution of Riemann problems at zone interfaces and a
final evolution stage.


\section{Results}

\subsection{Experiments/numerical simulations comparisons}

\begin{figure}[ht!]
\centering
\includegraphics[width=\columnwidth]{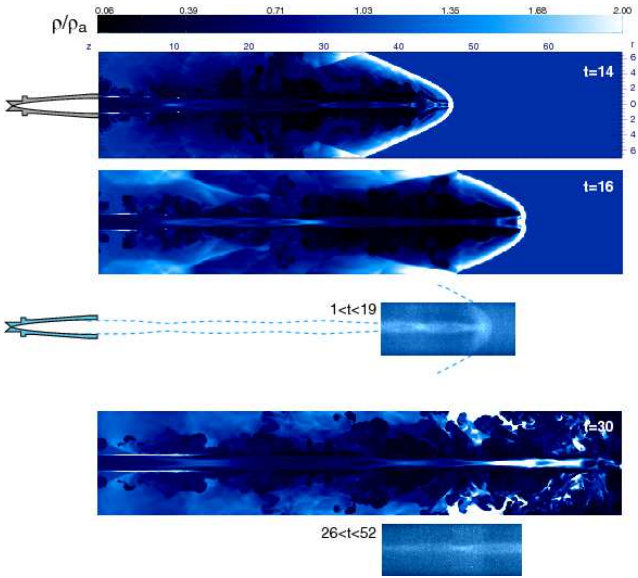}
\caption{Light jet, Helium in Xenon.
 $M$=16.1, ambient pressure $p_a = 4.0\pm 0.1$Pa, estimated nozzle exit velocity = 3200 $\pm$330 m/s,
mean jet pressure at nozzle exit $p_j = p_a \pm 30$\%, stagnation/ambient pressure ratio $p_0/p_a=7 \cdot10^4 \pm 30$\%.
Comparison  of numerical simulations (density maps) and visualizations (superpositions of scaled correlated frames).
The density maps are normalized to the unperturbed ambient value.
The space unit is the exit radius of the nozzle $r_0=0.03568$ m. The time unit for this light jet is $\tau=0.18$ ms}
\label{hexe}
\end{figure}

A slightly underdense (light) jet, Mach 15, density ratio 0.7 is shown in fig. \ref{hexe}. 

In this jet, dense knots form spontaneously on the axis well before the impact to the vessel end.
They are visible both in the numerical and in the experimental images. After the impact, 
the interaction with the reflected flow does not spoil the axial symmetry
of the flow. The impact of this jet on the vessel end wall reflects only a very 
small amount of matter, hence the perturbation may be considered weak. 
It is also noticeable that the inner part of the flow remains compact and collimated after the impact. 
This is also an interesting information, which could open the way to other applications, 
in particular to perturbative near-linear experimental studies. 
The inertial effects, and the associated compressibility, must here be so
powerful to inhibit the spatial growth of the jet, unlike what happens in the incompressible
situation, where sheared flows generally grow fast in thickness.

\begin{figure}[ht!]
\centering
\includegraphics[width=\columnwidth]{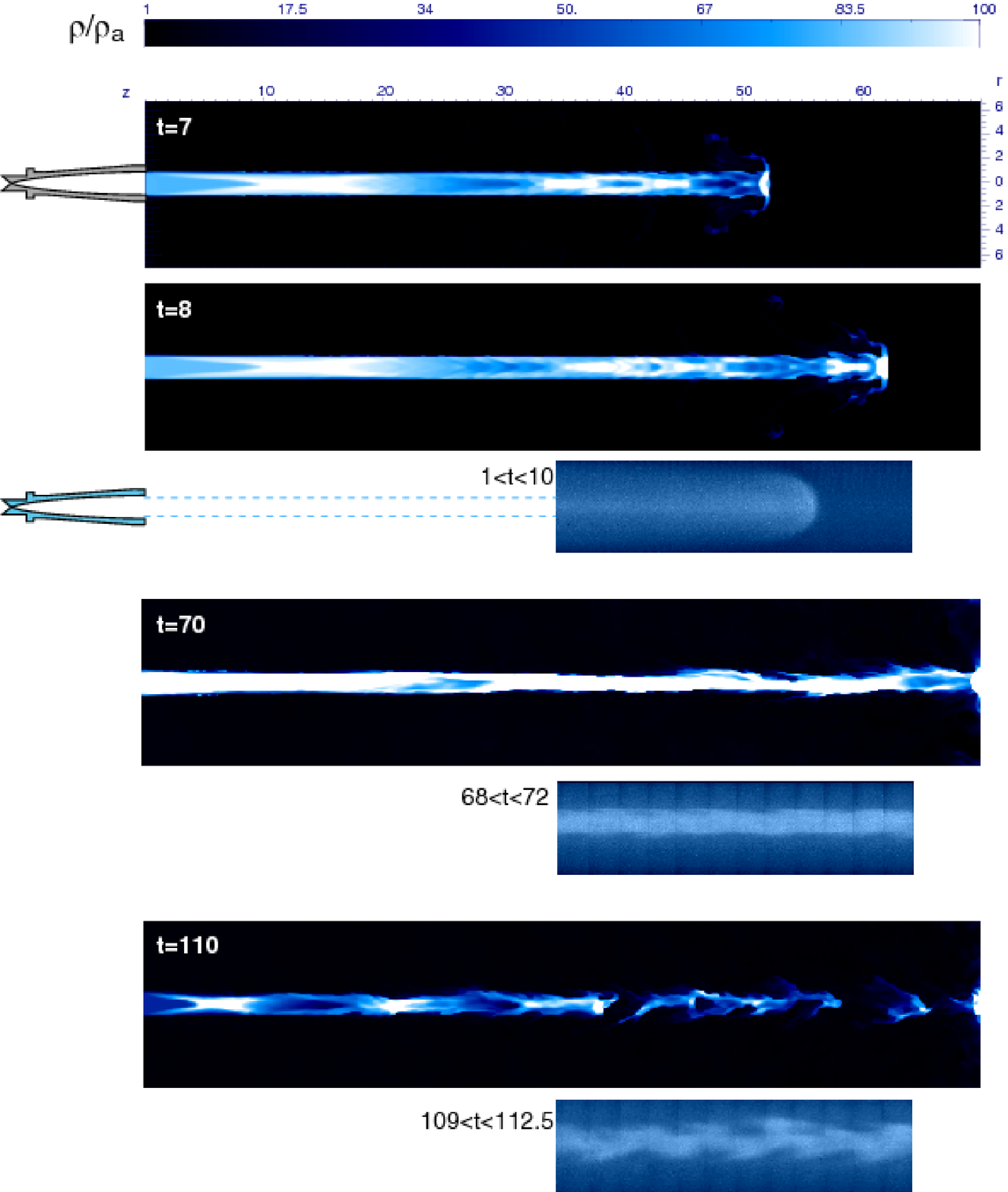}
\caption{Heavy jet, Xenon in Air.
$M$=15, ambient pressure $p_a = 9.95 \pm 0.1$ Pa, estimated nozzle velocity at the exit = 560 $\pm$ 60m/s,
mean jet pressure at the nozzle exit $p_j = p_a \pm 30$\%, stagnation/ambient pressure ratio $p_0/p_a=4.76 \cdot10^4 \pm 30$\%.
Comparison  of numerical simulations (density maps) and visualizations
(superpositions of scaled correlated frames).
The density maps are normalized to the unperturbed ambient value.
The space unit is the exit radius of the nozzle $r_0=0.03568$ m. The time unit for this heavy jet is $\tau=0.96$ ms}
\label{xeair}
\end{figure}

An overdense (heavy) jet, Mach 15, density ratio 100, is shown in fig. \ref{xeair}

Here, there are overdense zones on the axis, they are very close to each other, and are resolved only in the numerical simulation.
For the heavy jet, the interaction with the reflected flow eventually produces a finite perturbation, which non
linearly interacts with the outflow. This is sufficient to spoil the axial symmetry. 
However, it is not sufficient to spoil the collimation of the jet. This suggests the possibility of a very
large longitudinal extent of the heavy hypersonic jet and its related capability to transfer
energy, momentum and mass to large distances (on the Earth, in space, in possible new applications).

\begin{figure}[ht!]
\centering
\includegraphics[width=\columnwidth]{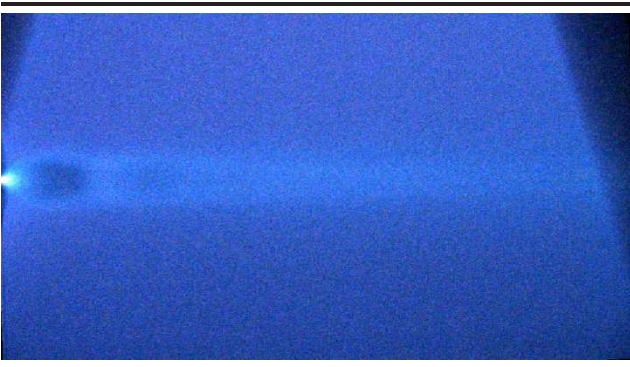}
\caption{Helium jet in an Argon medium: highly underexpanded jet, maximum Mach number = 16, stagnation/ambient pressure ratio 
$p_0/p_a = 304$, density ratio after the normal shock $\sim 0.15$.
}
\label{underexp}
\end{figure}

An example of hypersonic jet with a completely different behaviour is reported in fig. \ref{underexp}, 
where an underdense jet having Mach number and density ratio similar to the ones of the jet in fig. \ref{hexe} is shown (\cite{expf,pre}).
In this case, the jet is created by means of a truncated nozzle, and is  far from being nearly isoentropic. It is in fact characterized by a strong normal shock close to the nozzle exit, and by a lateral spreading which in the long term
is remarkably larger than that in fig. \ref{hexe}. The reduction of the core 
diameter is clear (the jet head is outside the image on the right) and is also a measure of the spreading of the annular mixing layer that surrounds it. The mixing region is slow compared to the flow in the  core and is not visible in this figure. However, it can be observed through  concentration measurements (\cite{expf,pre}). In this example the spreading angle with respect to the jet axis is about 8 degrees (\cite{pre}).

\subsection{Results analysis}

The velocities of flow structures (jet's head, knots, traveling waves..) can be measured by means of standard correlation
techniques applied to the pixel matrices of consecutive images. 

For the light jet of fig. \rf{hexe}, in the simulations the head (bow shock) velocity  grows from  $440\pm 3$m/s to $706\pm 3$m/s in the $2<t<16$ time interval (before the impact to the vessel end) due to the increase in the mass flow. The experimental head velocity at $t=15$ measured by image correlation is $490\pm 140$m/s.
For the heavy jet of fig. \rf{xeair}, the jet head velocity grows from $130\pm 3$m/s to $270\pm 3$m/s in the  $2<t<9$ time interval, 
and the experimental value at $t=8$ is  $140\pm30$m/s.

When the velocity of a structure is known, two images in succession containing the same structure can also be superimposed after shifting the second one by the displacement of the structure under study. Then, this procedure can be repeated, and a spatial reconstruction of a part of the jet structure
larger than the camera window can be done by superimposing parts of adjacent frames ('slices'). Clearly, the reconstruction of jet morphology on large scales has a physical meaning only if the changes in structure properties are slow with respect to the interframe time. 

\begin{figure}[h]
\centering
\includegraphics[width=\columnwidth]{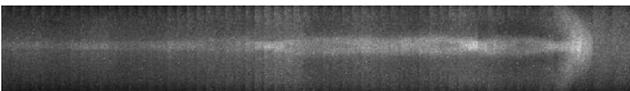}
\caption{'Frozen' image of the light jet analyzed in figure \ref{hexe}, Helium in Xenon at $M$=16.1, 
obtained by superimposed frames in the time range $0<t<66 \tau$} 
\label{corr1}
\end{figure}

\begin{figure}[h]
\centering
\includegraphics[width=\columnwidth]{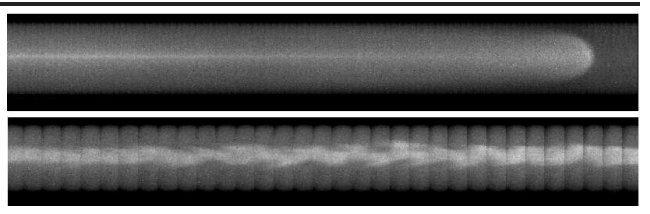}
\caption{'Frozen' images of a heavy jet, Xe in Ar at $M$=13.5, 
obtained by superimposed frames in the time ranges $0<t<20 \tau$ and $60\tau<t<70\tau$} 
\label{corr2}
\end{figure}

Two image reconstructions for a light jet and a heavy jet, obtained by correlated frames superposition, are shown in figures \ref{corr1} and \ref{corr2}.
These figures are constructed overcoming the validity range of the technique in order to obtain single pictures visualizing a whole jet or 
a large part of it, the resulting images must be considered as 'frozen', as if the jet were passing through the camera window without changes in its morphology.

\begin{figure}[h]
\centering
\includegraphics[width=\columnwidth]{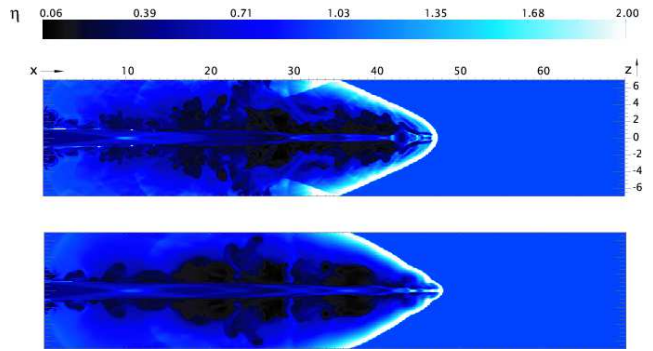}
\caption{Comparison between simulations with and without lateral boundaries at $t=14$}
\label{cut}
\end{figure}

The effect of the lateral boundaries of the vessel can be estimated by changing the
boundary conditions in the numerical simulations.
Figure \ref{cut} compares longitudinal 2D cuts, at $t=14$, of the light jet simulation obtained 
employing the actual lateral boundary conditions with the one obtained with outflow conditions, 
Differences in the density distribution morphologies are visible at the jet's head, that appears slightly narrower in 
the free-boundary simulation, while the head's positions differ by a few percent in the two cases.
The shock reflection at $x\sim 36$ is absent in the free-boundary case. 

\section{Conclusions}

Our experiment highlights the following aspects:

I) -  the collimation of near-isoentropic jets over distances much larger that the initial radius.

II) - the jet persistence (which, for the light jet, is mantained even after the impact 
with the vessel end-wall).

III) - the presence of a cocoon surrounding the underdense jet, while this is not visible
for the overdense one.

IV) - the spontaneous formation of knots along the jet axis (before the impact of the jet to the vessel end).

These aspects refer both to numerical simulations and to experiments ruled by Newtonian dynamics,
and do not need of the confining effect associated to the presence  of magnetic fields.

Some of these aspects are common to stellar jets, in particular the jet confinement and persistence over
long distances, in units of the initial radius.



\begin{thebibliography}{}


\bibitem[\protect\citeauthoryear{Ferrari}{1998}]{ferr98}
Ferrari, A.:
ARA\&A, {\bf 36}, 539 (1998) 

\bibitem[\protect\citeauthoryear{Reipurth and Bally}{2001}]{rb01}
Reipurth, B. and Bally, J., 
ARA\&A, {\bf 39}, 403 (2001)

\bibitem[\protect\citeauthoryear{Belan et al.}{2004}]{apss}
Belan M., De Ponte S., Massaglia S., Tordella D.:
\apss, {\bf 293} (1-2), 225-232 (2004) 

\bibitem[\protect\citeauthoryear{Belan et al.}{2006}]{kth}
Belan M., De Ponte S., Tordella D.:
{EFMC6 KTH, Euromech Fluid Mechanics Conference 6},
Royal Institute of technology, Stockholm (2006)

\bibitem[\protect\citeauthoryear{Belan, De Ponte and Tordella}{2008}]{expf}
Belan M., De Ponte S., Tordella D.:
{Exp Fluids, {\bf 45}, 3 (2008)}

\bibitem[\protect\citeauthoryear{Belan, De Ponte and Tordella}{2010}]{pre}
Belan M., De Ponte S., Tordella D.:
{Phys Review E, {\bf 82}, 2 (2010)}

\bibitem[\protect\citeauthoryear{Mignone et al.}{2007}]{mig07}
Mignone, A. et al.: 
ApJS, {\bf 170}, 228 (2007)

\bibitem[\protect\citeauthoryear{Mignone}{2009}]{mig09}
Mignone, A.:
Nuovo Cimento della Societa' Italiana di Fisica C-Colloquia on Physics, {\bf 32}, 2, 37-40 (2009)



\end{thebibliography}
\end{document}